\documentclass[paper]{JHEP3}
\usepackage{amssymb,enumerate}
\usepackage{amsmath}
\usepackage{bbm}
\usepackage{url}
\usepackage{cite}
\usepackage{graphics}
\usepackage{xspace}
\usepackage{epsfig}
\usepackage{subfigure}
\newcommand\POWHEG{{\tt POWHEG}}
\newcommand\POWHEGBOX{{\tt POWHEG BOX}}
\newcommand\POWHEGBOXVT{{\tt POWHEG-BOX-V2}}
\newcommand\PYTHIA{{\tt PYTHIA}}

\newcommand\HERWIG{{\tt HERWIG}}

\def\({\left(} 
\def\){\right)} 

\def\beq{\begin{equation}}
\def\beqn{\begin{eqnarray}}
\def\eeq{\end{equation}}
\def\eeqn{\end{eqnarray}}

\title{$W^+W^-$, $WZ$ and $ZZ$ production in the \POWHEGBOXVT{}}
\vfill
\author{Paolo Nason \\
INFN, Sezione di Milano Bicocca, Italy\\
E-mail: \email{Paolo.Nason@mib.infn.it}}
\author{Giulia Zanderighi \\
Rudolf Peierls Centre for Theoretical Physics, 1 Keble Road, University of Oxford, UK\\
E-mail: \email{g.zanderighi1@physics.ox.ac.uk}}


\abstract{We present an implementation of the vector boson pair
  production processes $ZZ$, $W^+W^-$ and $WZ$ within the \POWHEGBOXVT{}.
  This implementation, derived from the \POWHEGBOX{} version, has several
  improvements over the old one, among which the inclusion of
  all decay modes of the vector bosons, the possibility to generate
  different decay modes in the same run, speed optimization and phase
  space improvements in the handling of interference and singly resonant
  contributions.
}

\begin{document}
\section{Introduction}
In ref.~\cite{Melia:2011tj} an implementation of vector boson pair
production at NLO in QCD, that can be interfaced to a shower generator
according to the \POWHEG{} method, was presented. Only decays into
leptons and neutrinos were considered. Singly resonant contributions
and interference effects were included, and in fact all diagrams
relevant to the production of four leptons were accounted for, with
the exception of interference effects between $W^+W^-$ and $ZZ$
production when the decay products are the same. These last effects
were shown to be fully negligible at Born level.

In the present work, we present a new implementation of these
processes that has the following improvements over the old one:
\begin{itemize}
\item all possible decay modes are allowed;
\item different decay modes can be produced in a single run;
\item there is a considerable speed improvement;
\item there is an improvement in the treatment of the phase space and
  interference terms.
\end{itemize}

Although hadronic final states are all allowed, no NLO corrections to
the vector boson decays are included.
$W$ and $Z$ decays are in fact properly handled by
the shower Monte Carlo programs, like \PYTHIA{}
\cite{Sjostrand:2006za,Sjostrand:2007gs} and \HERWIG{}
\cite{Corcella:2000bw,Bahr:2008pv}, that can be interfaced to the
\POWHEGBOX{}, their resonance decay machinery being tuned to closely
reproduce collider data.

The new implementation exploits the fact that in the \POWHEGBOXVT{}
version one can specify if final state particles arise from a
resonance decay.  The algorithm for finding the radiation region in
real graphs, checks if a parton arises from a resonance rather than
from the production vertex, and handles the radiation accordingly. In
the present implementation, where no radiative corrections to decaying
resonances are considered, the only singular regions that are produced
by \POWHEG{} are thus relative to the production vertex.

\section{Matrix Elements}
We have used the same MCFM matrix elements \cite{Campbell:1999ah}
as in the old implementation, extending them to deal
with hadronic decays. A considerable increase in performance was achieved by storing intermediate
results in the matrix element calculation. Strong corrections in hadronic decays, of the
form $1+\alpha_{\scriptscriptstyle S}(M_V)/\pi$, are also included.

A non diagonal Cabibbo matrix is used by default (an arbitrary
CKM matrix can be entered in input by the user).  This is done in the
following way.  The only process in which flavour changing
interactions can arise in production is $WZ$ (in fact, in $W^+
W^-$ production flavour changing interactions are suppressed by the
GIM mechanism). In the $WZ$ case we thus generate explicitly from the
beginning all CKM allowed flavour changing processes.  In $W$ decays,
one generates the matrix elements for the decay into $\bar{u} d'$,
$\bar{c} s'$ (and the corresponding ones for the $W^+$), where $d'$
and $s'$ are the electroweak flavour eigenstates. Once the event is
generated, the $d'$ or $s'$ are transformed randomly into a $d$ or $s$
mass eigenstate, with a probability corresponding to the square of the
appropriate CKM entry.

\section{Phase space}
The treatment of the phase space, especially when handling processes with interference, has
been changed with respect to the original version.

When interference is present, we compute both the amplitudes $\cal{A}$ and $\cal{A}_{\rm exch}$,
where $\cal{A}_{\rm exch}$ is the amplitude with the final state identical particles exchanged.
In the previous version we computed the cross section using the squared amplitude
\begin{equation}\label{eq:asq}
 |{\cal A}|^2+|{\cal A}_{\rm exch}|^2+2{\rm Re}({\cal A}{\cal A}_{\rm exch}^*)\,.
\end{equation}
This had the disadvantage that the regions for the importance sampling of the resonances were
mixed up, and had to be handled with care~\cite{Melia:2011tj}.
In the present version, we instead do the following. We rewrite eq.~(\ref{eq:asq})
as
\begin{equation}\label{eq:asq1}
 \left(|{\cal A}|^2+|{\cal A}_{\rm exch}|^2\right)\times\left\{1+\frac{2{\rm Re}({\cal A}{\cal A}_{\rm exch}^*)}{|{\cal A}|^2+|{\cal A}_{\rm exch}|^2}\right\}\;.
\end{equation}
Noticing that both factors are symmetric for the exchange of the momenta of the
identical particles, and that $|{\cal A}|^2$ and $|{\cal A}_{\rm exch}|^2$ go
into each other by this exchange, we can as well use the expression
\begin{equation}
2 |{\cal A}|^2\times\left\{1+\frac{2{\rm Re}({\cal A}{\cal A}_{\rm exch}^*)}{|{\cal A}|^2+|{\cal A}_{\rm exch}|^2}\right\}   \,,
\end{equation}
that yields the same result upon integration. We then assign the resonances
according to the structure of the $|{\cal A}|^2$ term. The ambiguity in the assignment
is only present now in the small interference term, which is ignored, and no
particular importance sampling tricks are needed to handle it.

Notice that what would seem to be the most obvious choice
\begin{equation}
2 |{\cal A}|^2+2{\rm Re}({\cal A}{\cal A}_{\rm exch}^*)  \,.
\end{equation}
is in fact problematic, since it does not yield a positive definite cross section.

A further improvement was given by generating the partonic $s$ variable in the underlying Born
configuration with Lorenzian importance sampling over the possible single-resonant contribution. This
leads to a better description of the single-resonant region.

\section{Generation of the subprocesses}
In this version of the $VV$ production codes it is possible to generate the matrix elements for
all possible decays of the vector bosons. This leads to a large proliferation of amplitudes.
For example, there are 880 Born parton level configurations that arise in $WZ$ production.
Generating all processes at once slows down the program considerably. Furthermore, even if
the total result is computed with a satisfactory accuracy, it is not easy to check that
all decay processes have been accurately probed. This is particularly critical when computing
the upper bounds for radiation, since they are computed individually
for each underlying Born configuration. Because of this reason a new feature was introduced in
\POWHEGBOXVT{}, such that the upper bound for radiation of equivalent amplitudes
(i.e. amplitudes that are equal up to constant couplings) are combined. This feature is
activated by default for the processes at hand, but can also be activated for other
processes by including the line {\tt evenmaxrat 1} in the {\tt powheg.input} file.
Still it is often convenient to restrict the generated decay modes to the ones
one is really interested into.  All generators offer a number of
pre-defined options to select specific decay modes (e.g. leptonic, semi-leptonic, hadronic, etc.).
It is however difficult to anticipate all interesting possibilities. For this reason,
the code has been written in such a way that
the selection of decay modes can also be carried out by the user by
editing the subroutine {\tt alloweddec} in the {\tt init\_processes.f}
file. Further explanations on how to do this are given in the
respective manual.

\section{Availability}
The new code has been made available in the \POWHEGBOX{} svn repository,
and instructions for downloading the V2 version of \POWHEG{}
and the corresponding user processes are given at the URL
\url{http://powhegbox.mib.infn.it}. 

If you use this code, please quote the present note, and ref.~\cite{Melia:2011tj}.
Furthermore we remind that the matrix elements were obtained
from ref.~\cite{Campbell:1999ah}, and the \POWHEGBOX{} framework has been developed
in the sequel of publications \cite{Nason:2004rx,Frixione:2007vw,Alioli:2010xd}.


\providecommand{\href}[2]{#2}\begingroup\raggedright\endgroup

\end{document}